\documentclass[12pt,preprint]{aastex}

\slugcomment{Submitted to Astrophyisical Journal Letter}

\shorttitle{}
\shortauthors{Takami et al.}

\begin{document}


\title{A Micro Molecular Bipolar Outflow From HL Tau\footnote{
Based on observations obtained at the Gemini Observatory, which is
operated by the Association of Universities for Research in Astronomy,
Inc., under a cooperative agreement with the NSF on behalf of the
Gemini partnership: the National Science Foundation (United States),
the Particle Physics and Astronomy Research Council (United Kingdom),
the National Research Council (Canada), CONICYT (Chile), the Australian
Research Council (Australia), CNPq (Brazil) and CONICET (Argentina).}}


\author{Michihiro Takami\altaffilmark{2,3},
        Tracy L. Beck\altaffilmark{4},
        Tae-Soo Pyo\altaffilmark{2},
        Peter McGregor\altaffilmark{5},
        Christopher Davis\altaffilmark{6}
        }
\email{mtakami@subaru.naoj.org}


\altaffiltext{2}{Subaru Telescope, 650 North A'ohoku Place, Hilo,
HI, USA}

\altaffiltext{3}{Present Address: Institute of Astronomy and
Astrophysics, Academia Sinica, P.O. Box 23-141, Taipei 10617,
Taiwan, R.O.C.}

\altaffiltext{4}{Gemini Observatory, 670 North A'ohoku Place,
Hilo, HI 96720, USA}

\altaffiltext{5}{Research School of Astronomy and Astrophysics,
Australian National University, Cotter Road, Weston Creek, ACT
2611, Australia}


\altaffiltext{6}{Joint Astronomy Centre, 660 North A'ohoku Place,
University Park, Hilo, HI 96720, USA}


\begin{abstract}
We present detailed geometry and kinematics of the inner outflow
toward HL Tau observed using Near Infrared Integral Field
Spectograph (NIFS) at the Gemini-North 8-m Observatory. We
analyzed H$_2$ 2.122 $\micron$ emission and [\ion{Fe}{2}] 1.644
$\micron$ line emission as well as the adjacent continuum observed
at a $<$0\farcs2 resolution. The H$_2$ emission shows (1) a
bubble-like geometry to the northeast of the star, as briefly
reported in the previous paper, and (2) faint emission in the
southwest counterflow, which has been revealed through careful
analysis. The emission on both sides of the star show an arc
1\farcs0 away from the star, exhibiting a bipolar symmetry.
Different brightness and morphologies in the northeast and
southwest flows are attributed to absorption and obscuration of
the latter by a flattened envelope and a circumstellar disk. The
H$_2$ emission shows a remarkably different morphology from the
collimated jet seen in [\ion{Fe}{2}] emission. The positions of
some features coincide with scattering continuum, indicating that
these are associated with cavities in the dusty envelope. Such
properties are similar to millimeter CO outflows, although the
spatial scale of the H$_2$ outflow in our image ($\sim$150 AU) is
strikingly smaller than the mm outflows, which often extend over
1000--10000 AU scales. The position-velocity diagram of the H$_2$
and [\ion{Fe}{2}] emission do not show any evidence for kinematic
interaction between these flows. All results described above
support the scenario that the jet is surrounded by an unseen
wide-angled wind, which interacts with the ambient gas and produce
the bipolar cavity and shocked H$_2$ emission.

\end{abstract}





\section{Introduction}
Jets or outflows are associated with young stellar objects (YSOs)
at a variety of masses and evolutionary stages. There is growing
evidence that these are powered by mass accretion, although the
details of the flow launching mechanism are not yet clear.
Understanding this issue is hampered by the angular resolution of
the telescopes and interferometers, which is not yet high enough
to resolve the central engine of the driving source. Instead,
investigating flow geometry and kinematics close to the driving
source provides useful information for understanding the relation
between stellar mass accretion and outflows.

In particular, recent observations of near-infrared H$_2$ emission
have provided new clues to understand the activities of mass
ejection close to ($<$300 AU) low-mass protostars. Pioneering work
includes echelle spectroscopy and Fabry-Perot imaging of several
Class I protostars by Davis et. al. (2001, 2002). Davis et al.
(2001) revealed the presence of two blueshifted components at high
(50--150 km s$^{-1}$) and low velocities (5--20 km s$^{-1}$). Such
a kinematic structure in H$_2$ emission is similar to the
forbidden emission line regions close to T Tauri stars (see e.g.,
Hartigan et al. 1995). While the high-velocity component is
clearly associated with a collimated jet, the nature of the low
velocity component is less clear. This low velocity component
could be either a magnetohydrodynamically driven gas
 (e.g., Pyo et
al. 2003, 2004; Takami et al. 2004) or a component entrained by an
unseen wide-angled wind (e.g., Pyo et al. 2006; Takami et al.
2006).
Davis et
al. (2002) revealed H$_2$ emission associated with a collimated
jet and cavity walls toward some YSOs. Such emission is excited by
shocks at a temperature of $\sim$2000 $K$ (Takami et al. 2004,
2006; Beck et al. 2007).


Beck et al. (2007, hereafter Paper I) present the morphology and
excitation of the near-infrared H$_2$ emission toward six T Tauri
stars using NIFS (Near Infrared Integral Field Spectograph) at the
Gemini North Telescope on Mauna Kea. NIFS is an image slicing IFU
fed by Gemini's near-infrared adaptive optics system, Altair, to
obtain integral field spectroscopy with a resolving power of $R
\sim 5000$. These observations revealed a variety of morphological
distribution within 200 AU of T Tauri stars, and showed that H$_2$
molecules are presumably excited by shocks. The study we present
shows a detailed analysis of the emission associated with one of
the best studied YSOs, HL Tau. The YSO is in transient phase from
a Class I protostar to T Tauri star (see Pyo et al. 2006, and
references therein), and is known to host an extended collimated
jet (Mundt et al. 1990), millimeter CO outflow (Monin et al. 1996;
Cabrit et al. 1996), circumstellar disk (e.g., Wilner et al. 1996;
Kitamura et al. 2002), and also a flattened gas envelope (Sargent
\& Beckwith 1990) which may be infalling toward the star (Hayashi
et al. 1993). Paper I revealed the presence of a bubble-like
morphology in H$_2$ emission extending toward the northeast.

In this paper we analyze the results for H$_2$ 1-0 S(1) emission
(2.122 $\micron$) together with the [\ion{Fe}{2}] 1.644 $\micron$
line and also adjacent continuum.
Our results suggest the presence of an unseen wide-angled wind,
interacting with the ambient gas and produce a bipolar cavity and
shocked H$_2$ emission. Throughout the paper, we adopt the
distance to the target of 140 pc (Elias 1978).

\section{Observations and Results}

Observations were made using NIFS at the Gemini North Telescope on
February 8 and 11 2006. NIFS provides integral field spectroscopy
with a resolving power of $R=5300$ at 2.2 $\micron$. The NIFS
field is 3"$\times$3" in size and the individual IFU pixels are
0\farcs1$\times$0\farcs04 on the sky. The standard $K$- and
$H$-band settings provided the spectral ranges of 2.00--2.45
$\micron$ and 1.48--1.80 $\micron$, respectively. This allowed us
to observe H$_2$ 1-0 S(1) 2.122 $\micron$ and [\ion{Fe}{2}] 1.644
$\micron$ lines together with the adjacent continuum. Our spectra
also cover several other H$_2$ (see Paper I) and [\ion{Fe}{2}]
lines. The star was occulted using a 0\farcs2-$\phi$ mask to
provide long exposures (900 s for both $K$- and $H$-bands),
thereby observing the target lines with high signal-to-noise. The
total on-source integration were 3600 s for each band. In addition
to the target, a nearby sky field was observed for the background
subtraction, and A0 stars (HIP 15760 and HIP 25736) were observed
before and after the target for telluric correction. The Gemini
facility calibration unit, GCAL, was used with the IR continuum
($K$-band) and Quartz Halogen ($H$-band) lamps to obtain
flatfields (see Paper I for details).

Observations of HL Tau were acquired using the Gemini Adaptive
Optics system, Altair,using the nearby bright star XZ Tau for
wavefront reference ($\sim$ 20" distant). The ambient weather was
stable at 0\farcs5--0\farcs6 natural seeing conditions, but the
data were acquired through thin, sparse cirrus. The $H$-band and
$K$-band acquisition images of HL Tau showed that the central
source had an AO-corrected FWHM of 0\farcs17--0\farcs18. HL Tau
may not be a point source, hence these values provide upper limits
of the spatial resolution.

The raw IFU frames were reduced into datacubes using the NIFS
tasks in the Gemini IRAF package. After dark-subtraction, flat
fielding, field cosmetics, sky-background subtraction, 1-D
traditional long-slit spectra were extracted, and wavelength
calibration and telluric correction were made. Then these were
combined into 3-D data cubes ($x$, $y$, and $\lambda$). See Paper
I for more details.

Figure 1 shows the spatial distribution of H$_2$ 2.122 $\micron$,
[\ion{Fe}{2}] 1.644 $\micron$, and the adjacent continuum. As
briefly reported in Paper I, the H$_2$ emission shows a bright
bubble-like morphology in the northeast of the star. The angular
scale of this structure is 1\farcs2 and 0\farcs8 ($\sim$170 and
$\sim$110 AU) in length and width, respectively. The direction of
this feature roughly coincides with a blueshifted collimated jet
observed by Mundt et al. (1990) and Ray et al. (1996). The
structure of the bubble is in some extent axisymmetric about the
jet axis: the peak position at $\sim$0\farcs3 and $\sim$1\farcs0
away from the star is offset from the axis toward the right and
left, respectively. In addition to the northeast bubble, shown in
Paper I, we find a faint component seen in the opposite direction.
This southwest feature consists of (1) a relatively bright arc
$\sim$1\farcs0 away from the star, slightly elongated in the
east-west direction; and (2) a faint filamentary structure further
away toward the south. The position and morphology of the arc is
similar to the peak emission at the top of the northeast bubble.
These north-west and south-east emission structures represent the
blueshifted and redshifted flows, respectively, as shown later in
detail.

The [\ion{Fe}{2}] 1.644 $\micron$ shows a collimated morphology
with an angular width of 0\farcs2--0\farcs3, the width basically
increases with distance, except the base of the northeast jet,
which shows a width of $\sim$0\farcs3. The morphology in the
northeast is similar to the H$\alpha$ emission observed using the
{\it Hubble Space Telescope} (Ray et al. 1996). The emission shows
peaks at 0\farcs4 and 0\farcs7 away from the star, and faint
emission extends further away. Analogous to the H$_2$ emission,
the southwest emission is much fainter than the northeast,
typically by a factor of 10. The jet in this side exhibits a peak
1\farcs2 away from the star, and fainter emission extends to
further distances. These results agree with the previous echelle
spectroscopy by Pyo et al. (2006). Their position-velocity diagram
confirms that these small-scale jets are the inner regions of the
flows that extend over an arcminute scale observed by Mundt et al.
(1990).
%

In Figure 1, both H$_2$ and [\ion{Fe}{2}] emission show
bipolarity, in both cases the emission in the southwest is fainter
than the northeast by a factor of $\sim$10. Furthermore, the
southwest emission is absent within 0\farcs7--1\farcs0 of the
star. These characteristics are attributed to absorption of the
southeast (redshifted) flow by a flattened envelope extending over
$\sim$1000 AU (e.g., Sargent \& Beckwith 1990; Hayashi et al.
1993) and an optically thick circumstellar disk with a radius of
$\sim$100 AU (see Pyo et al. 2006, and references therein).
Hayashi et al. (1993) measured the radius and mass of the envelope
of 1400 AU and 0.03 $M_\odot$, respectively. These values yield an
average column density of $\sim$3$\times$10$^{22}$ cm$^{-2}$
assuming an inclination angle of 45$^\circ$ (see Pyo et al. 2006,
and references therein), thereby causing $A_K$$\sim$3. This should
reduce the H$_2$ 2.122 $\mu$m flux from the southwest flow by a
factor of $\sim$15, agreeing with the observed flux ratio between
the northeast and southwest flows. The circumstellar disk is
optically much thicker: the radius and mass of the disk of
$\sim$100 AU and 0.05 $M_\odot$ (Wilner et al. 1996; Kitamura et
al. 2002) yield $A_K \gg 100$. This should thus completely obscure
the southwest flow at least within 0.5 AU of the star, as is
observed.

The continuum emission at 2.12 and 1.64 $\micron$ shows a triangular
morphology to the north-east due to scattering in a cavity in the dusty envelope
(e.g., Weintraub et al. 1995; Close et al. 1997; Lucas et al. 1994).
This nebulosity is more clearly seen at 1.64 $\micron$ presumably
due to the following two facts: (1) the flux from the star is more
affected by extinction, thereby relatively faint; (2) the scattering
efficiency is higher at shorter wavelengths.
The 1.64 $\micron$ continuum also shows a marginal extension
toward the west (see the bottom-right of the image in Figure 1).
Unlike the H$_2$ and [\ion{Fe}{2}] emission, the continuum
emission does not seem to show any other remarkable structures in
the southwest.

We perform detailed comparison of the H$_2$, [\ion{Fe}{2}] and
continuum images in Figure 2. Included in the bottom region of
Figure 2 are 1-D slices of the spacial flux distribution across
the axis of the jet. The grayscale H$_2$ images together with the
contour of the continuum emission show that the H$_2$ flows are
bracketed by a cavity in the dusty envelope. Indeed, the position
of the left side of the northeast bubble (position A in the
figure) matches well with the ridge in the continuum image. In the
1-D strips, the position of H$_2$ and [\ion{Fe}{2}] emission also
match each other at positions B and C.

Figure 3 shows an unsharp-masked image of the 1.64 $\micron$
continuum. For this image processing, the original image is
smoothed using a rectangular point-spread function with a size of
15$\times$1 pixel$^2$ (1\farcs2$\times$0\farcs08), and subtracted
from the original image. This enhances the morphology of the
ridges, which are seen in the profiles in Figure 2. Figure 3
clearly shows a V-shape morphology in the northeast with an
opening angle of $\sim$60$^\circ$. The axis of the V-shape
structure matches well with the jet axis seen in the [\ion{Fe}{2}]
image in Figure 1. Although the possible bias introduced by
unsharp masking may not be clear, we believe this structure in
Figure 3 outlines the shape of the inner cavity walls. As shown by
Figure 1 and also previous observations (e.g., Lucas et al. 2004),
the brightness distribution of the continuum shows asymmetry about
the flow/cavity axis. The northeast emission is much brighter at
the left side of the axis. Similar asymmetry in the scattering
nebulae is also seen toward some other YSOs (e.g., HH 30 ---
Cotera et al. 2001; HH 46/47 --- Heathcote et al. 1996), and this
could result from non-uniform brightness distribution of the
stellar surface due to the presence of hot spots (e.g., Cotera et
al. 2001). In the southwest, the scattered light continuum
emission is seen only in the bottom-right of the image in Figure
3, and the emission associated with the bottom-left side of the
cavity appears to be missing. Instead, the position of the faint
filamentary structure seen in the H$_2$ emission corresponds to
the inner cavity wall that is not seen in continuum emission
(overplotted to the right in Figure 3).

Figure 4 shows the position-velocity diagram of the H$_2$ and
[\ion{Fe}{2}] emission along the top-bottom direction of Figure 1.
The H$_2$ emission shows velocities nearly the same as the
systemic velocity, and the northeast flow appears slightly
blueshifted than the southwest flow (by $\sim$15 km s$^{-1}$).
Thus, it is likely that each side of the H$_2$ flow are moving
toward or away from us with a radial velocity of 5--10 km s$^{-1}$
compared with the systemic velocity. Such a velocity is comparable
to the expanding velocity of the cavity measured by low-order
adaptive optics imaging in different years ($\sim$30 km s$^{-1}$,
Close et al. 1997). The velocity of the H$_2$ outflow would be
remarkably larger than the molecular bipolar outflow extending
over $\sim$20" scales (1--1.5 km s$^{-1}$, Monin et al. 1996;
Cabrit et al. 1996).

Such velocities seen in the H$_2$ emission contrast with the
velocity of the collimated jet seen in the [\ion{Fe}{2}] emission.
In Figure 4, the [\ion{Fe}{2}] emission shows LSR
velocities of $\sim$--180 and $\sim$150 km s$^{-1}$ at the northeast
and southwest, respectively. These agree with the previous
measurements by echelle spectroscopy at Subaru (Pyo et al. 2006).
The figure shows that the velocity of the H$_2$ and [\ion{Fe}{2}]
emission is constant over each side of the star at a spectral
resolution of NIFS. Their line widths coincide with the resolving power
($\sim$70 km s$^{-1}$), revealing that the velocity dispersion of each
flow component is spectrally unresolved.

\section{Discussion}
Mass ejection from low-mass YSOs is often observed in two different
manners: (1) collimated jets often seen in optical-IR wavelengths, or
millimeter wavelengths in some cases
(see e.g., Bally et al. 2007 for review) (2) molecular bipolar
outflows, which show a variety of morphologies in millimeter CO lines
(see e.g., Arce et al. 2007 for review). The relation between these
types of flows remain unclear. In particular, two major scenarios
have been discussed over decades for driving of the molecular
outflows. These are: (1) jet-driven scenario, for which the
molecular outflow results from interaction between a collimated jet
and ambient material (e.g., Raga \& Cabrit 1993); and (2)
wind-driven scenario, for which the molecular outflow results from
an unseen wide-angled wind (e.g., Shu et al. 1991). Studies to date
suggest that neither the jet-driven nor wind-driven models can explain a
wide range of morphologies and kinematic properties observed in all
outflows (see Cabrit et al. 1997; Arce et al. 2007).

The observed morphology and kinematics in H$_2$ emission from HL
Tau are similar to millimeter CO outflows.
It is striking that the H$_2$ flow observed in HL Tau is smaller
than a typical spatial scale of the millimeter CO outflows. While
the CO outflows often extend over 1000--10000 AU scales (see e.g.,
Lee et al. 2000, 2002), including HL Tau itself (Monin et al.
1996; Cabrit et al. 1996), the H$_2$ flow in HL Tau shows a
spatial scale of only $\sim$150 AU.

The H$_2$ emission shows a remarkably different geometry and
kinematics from the [\ion{Fe}{2}] emission associated with the
collimated jet. Furthermore, there is no evidence for kinematic
interaction between the H$_2$ flow and the collimated jet: i.e.,
there is no clear evidence of acceleration of deceleration of the
H$_2$ and [Fe II] flows where these are overlapped in Figure 2.
These results support the scenario that the jet is surrounded by
an unseen wide-angled wind, which interacts with the ambient gas
and produces the bipolar cavity and also shocked H$_2$ emission.
The presence of such a wide-angled wind indeed agrees with
proposed magneto-hydrodynamically driven wind models (see e.g.,
Brandford \& Payne 1982; Uchida \& Shibata 1985; Shang et al.
2006). Alternatively, Close et al. (1997) suggests that the
cavities associated with HL Tau could be produced by a precessing
jet. However, we emphasize that our results do not show any
evidence for kinematic interaction between the [\ion{Fe}{2}]
emission associated with the collimated jet and the H$_2$ emission
at the cavity walls. Furthermore, a typical cooling time scale of
the near-infrared H$_2$ emission line regions is less than a year
(see e.g.,Takami et al. 2004), and the gas associated with the jet
would move only $<$0\farcs2 in such a period (Mundt et al. 1990).
Thus, the jet would have to be spatially located much closer to
the H$_2$ emission line regions if it was responsible for the
H$_2$ line excitation.

The arcs in H$_2$ emission 1''.0 away from the star suggests that
the ejection activity is time-variable. Assuming a flow velocity
of 10 km s$^{-1}$ inclination angle of 34$^\circ$ from the plane
of the sky (Mundt et al. 1990), we estimate the dynamical age of
these arcs of $\sim$70 yrs. Periodic monitoring observations
spread over year-long timescales are necessary to measure their
proposer motions and determine the accurate dynamical age. Such a
time-variable wind may be a rather common activity associated with
low-mass YSOs. Indeed, a wind bubble similar to the northwest
H$_2$ emission is also observed in the neighboring YSO XZ Tau in
optical emission lines (Krist et al. 1997, 1999; Coffey et al.
2004). Furthermore, some millimeter CO outflows show multiple
shell structures extending over 1000--10000 AU scales (see e.g.,
Lee et al. 2002). Arce \& Goodman (2001) show that episodic
outflows well explain the presence of ``Hubble wedges'' (i.e., a
jagged profile) in the position-velocity diagram, and also steep
power-low slopes of mass-velocity relation ($dM(v)/dv \propto
v^{-\gamma}$, $\gamma > 2$) observed toward some outflows.

Time-variable mass mass ejection may be related to episodic mass
accretion, observed in extreme cases, as FU Orionis (FUor) or EX Orionis (EXor)
outbursts (see e.g., Hartmann \& Kenyon 1996; Herbig 1989). Indeed, optical-IR absorption
lines of FUors suggest the presence of an energetic disk
wind (e.g., Calvet et al. 1993). Furthermore, spectroscopic monitoring
of CO overtone absorption suggests that one of the FUors ejected an expanding shell
of dense, low-temperature material within a few decades ago (Hartmann et al. 2004).
Such a link between episodic
mass ejection and accretion has also been proposed to explain the
morphology of the collimated jet associated with YSOs
(e.g., Dopita 1978; Reipurth 1989; Zinnecker et al. 1998).

\acknowledgments
We are very
grateful to the Gemini North NIFS SV Team, Queue Observers, and Systems
Support Associates for their assistance acquiring these observations;
particularly to Inseok Song, Gelys Trancho and Brian Walls.  Data for
this program were acquired at Gemini North Observatory during System
Verification of coronographic observations with the NIFS Integral Field
Spectrograph under Gemini program ID: GN-2006A-SV-129.
Some data analysis was made using the Starlink software developed by
CCLRC on behalf of PPARC.
This research has also been made use of the NASA's Astrophysics Data
System Abstract Service.



{\it Facilities:} \facility{Gemini (NIFS)}.



\appendix




\clearpage




\begin{figure*}
\epsscale{1.0}
\plotone{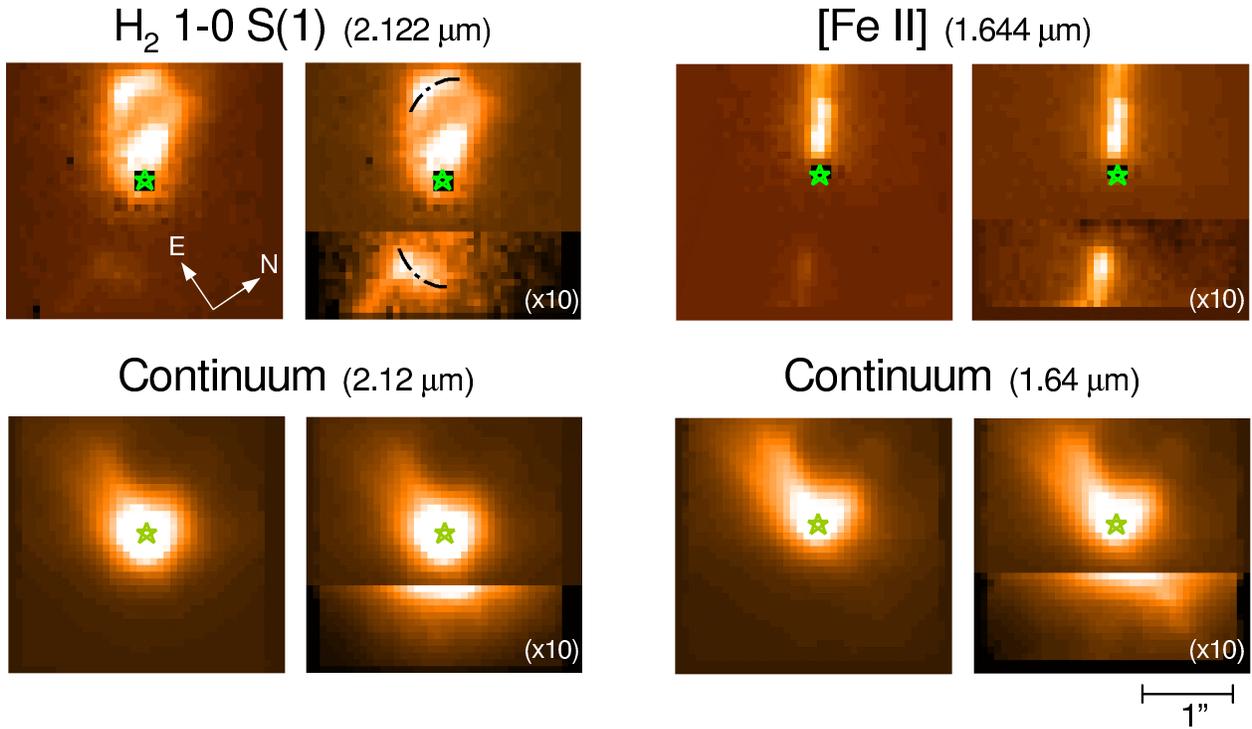} \caption{Continuum-subtracted images of H$_2$
2.122 $\micron$ ($V_{\rm LSR}$=$-$50 to 80 km s$^{-1}$),
[\ion{Fe}{2}] 1.644 $\micron$ ($V_{\rm LSR}$=$-$200 to $-$110 and
100 to 190 km s$^{-1}$), and adjacent continuum. The upper
direction corresponds to P.A.=56$^\circ$. In the right images the
flux of the bottom region is magnified by a factor of 10 to show
faint structures. The position of the marked stellar position is
determined by a peak of the continuum and/or the position of the
occulting mask. A symmetric structure seen in the H$_2$ image is
indicated by dot-dashed curves.\label{fig1}}
  \label{whole_spec}
\end{figure*}

\clearpage



\begin{figure}
\epsscale{0.5}
\plotone{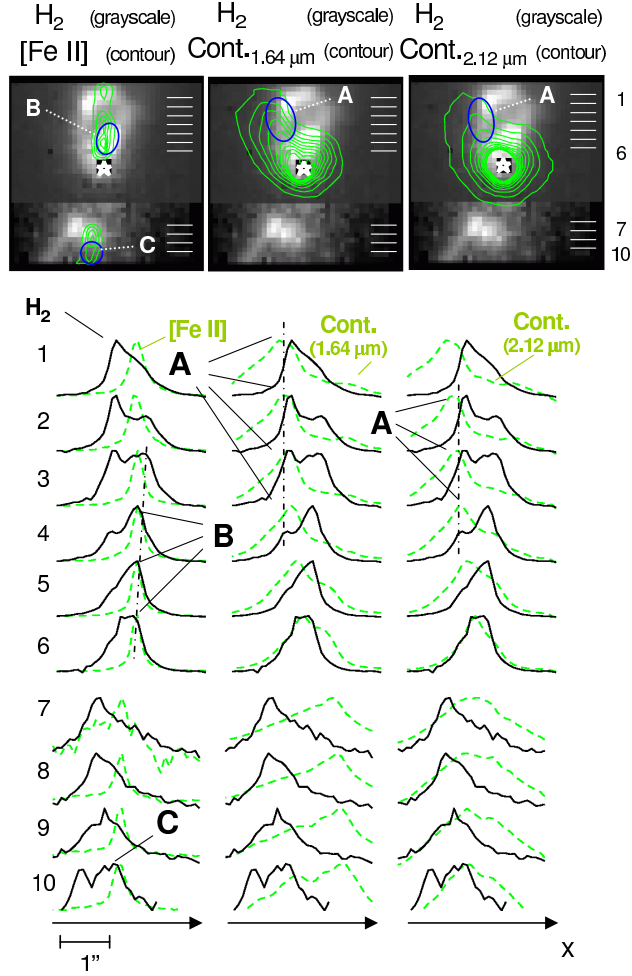}
\caption{(top) the H$_2$ 2.122 $\micron$ image with contour of
[\ion{Fe}{2}] emission, and also continuum at 1.64 and 2.12
$\micron$. The features discussed in the main text are labeled as
A, B, and C. The white ticks at the right show the positions where
we extracted 1-D intensity distribution for the plots at the
bottom. (bottom) Spatial cuts of the H$_2$, [Fe II] and continuum
flux (1.64/2.12 $\micron$) across the axis of the outflow. The
H$_2$ profiles drawn with full lines are compared to the [Fe II]
and continuum profiles shown in dashed lines. Each profile is
arbitrarily scaled.\label{fig2}}
\end{figure}

\begin{figure}
\epsscale{0.5}
\plotone{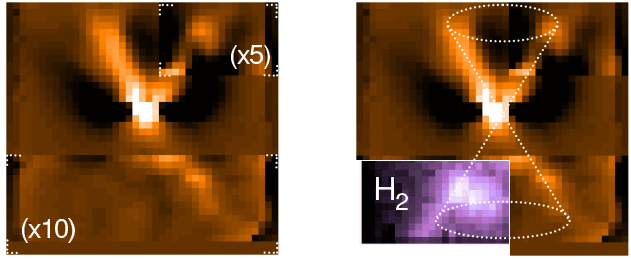} \caption{(left) unsharp masked image of the
continuum at 1.64 $\micron$. The original image is smoothed using
a rectangular point-spread function with a size of 15$\times$1
pixel$^2$ (1\farcs2$\times$0\farcs08), and subtract from the
original image. The flux is magnified by a factor of 5 and 10 in
the top-right and bottom regions, respectively, to clearly show
the structures detected. (right) same as the left figure, but the
H$_2$ 2.122 $\micron$ image is shown in the bottom-left. The
dotted cone shows a probable morphology of the outflow cavity.
\label{fig3}}
\end{figure}

\begin{figure}
\epsscale{0.5}
\plotone{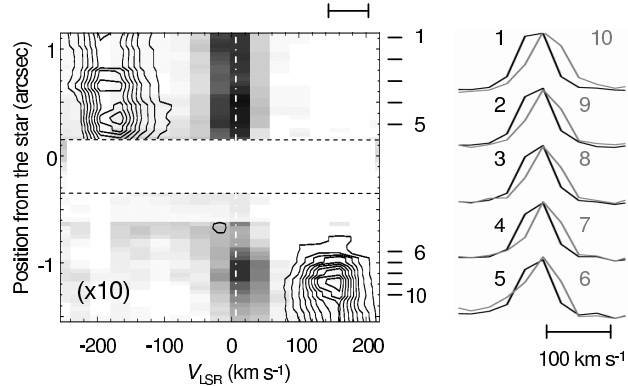}
\caption{(left) Position velocity diagram of the H$_2$ 2.122
$\micron$ (grayscale) and [\ion{Fe}{2}] 1.644 $\micron$ emission
(contour). The intensity is magnified by a factor of 10 in the
bottom half of the diagram. The contours are shown from 9.5 \% to
95 \% of the peak intensity with equal intervals in a linear
scale. The region close to the star is masked as the residual of
continuum subtraction is seen. The dot-dashed line shows the
systemic velocity of HL Tau measured from the molecular envelope
($V_{\rm LSR}$=+6 km s$^{-1}$, Hayashi et al. 1993). The scale bar
at the top-right shows the spectral resolution of the instrument.
(right) the line profiles of the H$_2$ emission at position 1--10,
extracted from the left figure. The solid and gray lines show the
profiles in the upper and bottom half of the PV diagram. The value
of the peak intensity is arbitrarily scaled. As seen in the
figure, the H$_2$ emission at the top half of the PV diagram is
slightly blueshifted compared with the bottom half.\label{fig4}}
\end{figure}







\clearpage






\end{document}